\documentclass[preprint]{aastex}

\received{2000 August 7}
\accepted{2000 September 27}

\slugcomment{To appear in ApJL.}

\shorttitle{NUMEROUS OLD STARBURST GALAXIES}
\shortauthors{MOURI \& TANIGUCHI}

\begin{document}

\title{NUMEROUS OLD STARBURST GALAXIES IN THE LOCAL UNIVERSE}

\author{HIDEAKI MOURI}
\affil{Meteorological Research Institute, Nagamine 1-1, Tsukuba 305-0052, Japan; hmouri@mri-jma.go.jp}

\and

\author{YOSHIAKI TANIGUCHI}
\affil{Astronomical Institute, Graduate School of Science, Tohoku University, Aoba, Sendai 980-8578, Japan}

\begin{abstract}

Old starburst galaxies are deficient in O stars, and hence do not exhibit strong line emission in the optical regime. However, there remain many B stars, which are expected to heat dust grains and generate strong continuum emission in the far-infrared. The {\it IRAS} data for a statistically complete sample of nearby galaxies reveal for the first time that such objects are as numerous as 30--40\% of the local galaxy population.

\end{abstract}


\section{INTRODUCTION}
Starburst galaxies (SBGs) are selected usually by optical emission lines. They originate in ionized gas around O stars. The classical examples are the Markarian SBGs, which account for 3\% of the field galaxies (Balzano 1983). However, the above selection is biased toward young SBGs. The Markarian SBGs are actually at starburst ages of $\le$ 20 Myr (Calzetti 1997). A starburst event generates many B stars, which live longer than O stars. With increasing the starburst age and decreasing the star formation rate, O stars become few but B stars remain numerous. Such old SBGs have not been explored, except for some case studies (Rieke, Lebofsky, \& Walker 1988; Koornneef 1993). In particular, their space density is unknown. The present knowledge of SBGs is thus incomplete.

By comparing the Markarian objects with SBGs in the Palomar Survey (Ho, Filippenko, \& Sargent 1997a, b, c), we propose that old SBGs outnumber classical SBGs by a factor of $\sim$10 and make up major constituents of the nearby galaxies. The Palomar Survey is an optical spectroscopic survey for nuclear regions on a magnitude-limited sample of 486 galaxies ($B_T$ $\le$ 12.5 mag), is more sensitive to faint line emission than previous surveys, and thus provides a fair representation of the local galaxy population. Roughly 40\% of the Palomar galaxies exhibit line emission, albeit weak, resembling those of \ion{H}{2} regions. We find that many of these Palomar SBGs are as luminous as the Markarian SBGs in the far-infrared (FIR). However, the Palomar SBGs are not so luminous in the mid-infrared (MIR). Since the FIR and MIR continua arise from dust grains heated by B and O stars, respectively (Mouri \& Taniguchi 1992), it is concluded that the Palomar SBGs are dominated by old objects where O stars are few. We confirm this conclusion with a model for infrared spectra of SBGs, and discuss some of its implications.

Throughout this Letter, the term SBG is used to represent a galaxy harboring a starburst region. Most of the Palomar and Markarian SBGs are galaxies with nuclear starbursts. This is because Ho et al. (1997a, b, c) and Balzano (1983) surveyed galaxy nuclei. We focus on SBGs dominated by starbursts, and study their total FIR and MIR emissions.

\section{OBSERVATIONAL DATA AND THEIR STATISTICS}
The present work is based on the FIR continuum fluxes at 60 and 100 $\mu$m as well as the MIR continuum flux at 25 $\mu$m, measured by the {\it Infrared Astronomical Satellite} ({\it IRAS}). For the Palomar SBGs, we use the data compiled by Ho et al. (1997a). For the Markarian SBGs, the data are from Soifer et al. (1989) and Moshir et al. (1990). These {\it IRAS} fluxes represent emissions from the whole galaxy. The {\it IRAS} apertures are large enough to encompass most of the galaxies. Spatially integrated fluxes are used for extended sources.

The statistical properties are summarized in Figure 1, where filled areas denote the Palomar SBGs and open areas denote the Markarian SBGs. Only a few objects are common to the two samples. The object numbers and median values are given in the figure caption. 


Figure 1$a$ shows the distribution of the FIR luminosity $L_{\rm FIR}$. The FIR flux between 40 and 120 $\mu$m is defined as $1.26 \times 10^{-14} (2.58 S_{60} + S_{100})$ W m$^{-2}$ (Fullmer \& Lonsdale 1989). Here $S_{\lambda}$ is the flux density at $\lambda$ $\mu$m in janskys. The galaxy distances are taken from Ho et al. (1997a), or calculated from the redshifts in the cosmic-microwave-background frame with $H_0$ = 75 km s$^{-1}$ Mpc$^{-1}$. The FIR luminosity exceeds 10$^{10}$ $L_{\odot}$ in 30\% of the Palomar SBGs and 50\% of the Markarian SBGs. Hereafter we focus on these SBGs. Their FIR emission is dominated by starbursts. The archetypical SBGs such as M82 and NGC 253 actually exhibit $L_{\rm FIR}$ $\simeq$ a few $\times$ 10$^{10}$ $L_{\odot}$.  For most of the Palomar and Markarian SBGs with $L_{\rm FIR}$ $>$ 10$^{10}$ $L_{\odot}$, we find $L_{\rm FIR}$ $>$ $L_{B_T}$. Here $L_{B_T}$ is the total $B$-band luminosity, and $L_{\rm FIR}$ $>$ $L_{B_T}$ is expected for a strong starburst (Moorwood 1996; Sanders \& Mirabel 1996).

Figures 1$b$ and 1$c$ show, respectively, the distributions of the spectral indices $\alpha (25, 60)$ and $\alpha (60, 100)$. They are defined as $\alpha (\lambda _1 , \lambda _2 )$ = $\log (S_{\lambda _1} / S_{\lambda _2} ) / \log (\lambda _2 / \lambda _1 )$. The Palomar SBGs prefer low values of $\alpha (25, 60)$ and $\alpha (60, 100)$, while the Markarian SBGs prefer high values. The probabilities that the two samples represent different distributions are $\ge$ 99.99\%, if we apply the generalized Wilcoxon tests for data with upper limits (Feigelson \& Nelson 1985) or the usual Kolmogorov-Smirnov test. When we plot $\alpha (25, 60)$ or $\alpha (60, 100)$ against $L_{\rm FIR}$, the Palomar and Markarian SBGs are separated at each of the values of $L_{\rm FIR}$. The Palomar SBGs with the lowest values of $\alpha (25, 60)$ tend to exhibit the lowest values of $\alpha (60, 100)$. Thus, although many of the Palomar SBGs have FIR luminosities being comparable to those of the Markarian SBGs, their {\it IRAS} colors are systematically different.

\section{MODEL ANALYSIS}

The difference in $\alpha (25, 60)$ between the Palomar and Markarian SBGs is due to a difference in the starburst age. To demonstrate this, we reproduce the $\alpha (25, 60)$ value with an empirical model (Mouri \& Taniguchi 1992), where two emission components with fixed colors are related to the ionizing-photon flux $N_{\rm ion}$ and the bolometric luminosity $L_{\rm bol}$. They are in turn calculated with evolutionary synthesis models. 

Our approach is supported by an anti-correlation between the spectral index $\alpha (25, 60)$ and the flux ratio [\ion{O}{1}] 6300\AA/H$\alpha$ among SBGs (Mouri \& Taniguchi 1992; Larkin et al. 1998). The [\ion{O}{1}]/H$\alpha$ ratio is a measure of the number ratio of supernovae to O stars. A variation in the $\alpha (25, 60)$ value reflects a variation in the stellar population rather than color variations in the individual emission components. The color variations, together with a variation in the excitation efficiency of the [\ion{O}{1}] emission, are merely responsible for scatter around the anti-correlation.

There are two types of dust grains. Large grains are in thermal equilibrium with the ambient radiation field, while small grains are heated transiently to high temperatures by single photons. Since small grains are destroyed in the vicinity of O stars, the infrared continuum is a mixture of i) MIR emission from hot large grains within ionized gas, ii) FIR emission from warm large grains outside ionized gas, and iii) MIR emission from small grains outside ionized gas (see Fig. 1 of Mouri, Kawara, \& Taniguchi 1997\footnote{
This paper has typographical errors in \S2. The {\it IRAS} apertures at 12 and 25 $\mu$m should be read as 0\farcm76 $\times$ 4\farcm6. The second equation of the footnote 1 should be read as $S_{\lambda}/F_{\rm LG}$ = $2.4 \times 10^8 \lambda ^{-5} T_{\rm d} ^{-6} [ \exp (14387/( \lambda T_{\rm d} )) -1]^{-1}$ Hz$^{-1}$.}). 
To indicate that a quantity is relevant to hot large, warm large, large, or small grains, we hereafter use the suffix HLG, WLG, LG, or SG.

We assume that grains absorb all the stellar photons, except for those used in ionizing the gas. Since some of nonionizing photons actually escape from a starburst region, our model overestimates the importance of those photons. Nevertheless, this effect is not serious. Starburst regions are quite dusty. For example, the amount of extinction derived from the Br$\alpha$/Br$\gamma$ ratio is much greater than that derived from the H$\alpha$/H$\beta$ ratio (Kawara, Nishida, \& Phillips 1989).

Hot large grains are heated by Ly$\alpha$ and ionizing photons. Their luminosity is given by $L_{\rm HLG}$ = $((1-f) \, h\nu _{{\rm Ly}\alpha} + f \, h\nu _{\rm ion}) N_{\rm ion}$. Here $f$ is the fraction of ionizing photons absorbed by the grains, $h\nu _{{\rm Ly}\alpha}$ is the energy of a Ly$\alpha$ photon, and $h\nu _{\rm ion}$ $\simeq$ $2 h\nu _{{\rm Ly}\alpha}$ is the mean energy of ionizing photons. Large grains radiate as a greybody with an emissivity being proportional to $\lambda ^{-1}$:
\begin{equation}
\label{a}
\frac{S_{\lambda}^{\rm LG}}{F_{\rm LG}} =
\frac{8.262 \times 10^4 \lambda ^{-4} T_{\rm LG}^{-5}}
     {\exp (14387/( \lambda T_{\rm LG} )) -1}
     \quad
     {\rm Hz}^{-1}.
\end{equation}
Here the wavelength $\lambda$ is in microns and the grain temperature $T_{\rm LG}$ is in kelvins. Mouri et al. (1997) estimated as $T_{\rm LG}$ = $T_{\rm HLG}$ $\simeq$ 140 K, from a comparison of the {\it IRAS} fluxes with the Br$\gamma$ flux on a sample of SBGs. Mouri et al. also found a relation $\nu _{25} S_{25}^{\rm HLG}$ = $3400 F_{{\rm Br}\gamma}$, which yields $f$ $\simeq$ 0.5 if the case B recombination (Osterbrock 1989) is assumed for the Br$\gamma$ emission. These values are used in our calculation. Baldwin et al. (1991) studied the Orion \ion{H}{2} region and obtained similar results for $T_{\rm HLG}$ and $f$. The {\it Infrared Space Observatory} detected MIR emission from hot large grains in galaxies (Vigroux et al. 1996; Dale et al. 2000; see also Dultzin-Hacyan, Masegosa, \& Moles 1990).

Warm large grains and small grains are heated by nonionizing photons. Their luminosities are given by $L_{\rm WLG}$ $+$ $L_{\rm SG}$ = $L_{\rm bol} - h \nu _{\rm ion} N_{\rm ion}$. We use the model of Boulanger et al. (1988), which is based on the {\it IRAS} observation of the California Nebula. Small grains absorb 45\% of the nonionizing photons. The emergent spectrum is $S_{12}^{\rm SG} : S_{25}^{\rm SG} : S_{60}^{\rm SG} : S_{100}^{\rm SG}$ = $1 : 1.1 : 1.7 : 2.2$ with $S_{12}^{\rm SG} / (F_{\rm WLG} + F_{\rm SG} )$ = $4 \times 10^{-15}$ Hz$^{-1}$. Although small grains emitting at 12 $\mu$m could be destroyed even outside ionized gas, this effect is negligible at $\ge$ 25 $\mu$m (Boulanger et al. 1988; Mouri et al. 1997). Warm large grains absorb the rest of the nonionizing photons and reradiate as the greybody (\ref{a}). Since the median values of $\alpha (60, 100)$ in the Palomar and Markarian SBGs correspond to $T_{\rm LG}$ = $T_{\rm WLG}$ $\simeq$ 30--40 K, we adopt $T_{\rm WLG}$ = 35 K. The change of $T_{\rm WLG}$ between 30 and 40 K is unimportant to the value of $\alpha (25, 60)$, as shown below in Figure 2$c$.

The quantities $N_{\rm ion}$ and $L_{\rm bol}$ are obtained as a function of the starburst age $t$ from solar-metallicity synthesis models of Leitherer et al. (1999). They adopted a Salpeter initial mass function with the lower and upper cutoffs of 1 and 100 $M_{\odot}$. We convolve their instantaneous burst models to obtain the values of $N_{\rm ion}$ and $L_{\rm bol}$ for star formation rates decaying exponentially as $\exp (-t/ \tau )$. The starburst duration $\tau$ is set to be 1, 2, 5, 10, 20, or 50 Myr. We also use their constant star formation models ($\tau$ = $\infty$). 


The results are summarized in Figure 2: ($a$) $N_{\rm ion}$, ($b$) $L_{\rm FIR}$, and ($c$) $\alpha (25, 60)$. The ionizing-photon flux $N_{\rm ion}$ and the FIR luminosity $L_{\rm FIR}$ are normalized by their maximum values. When the starburst is young, ionizing photons are abundant. Since the MIR emission from hot large grains is strong, the spectral index $\alpha (25, 60)$ is flat and close to the values observed in the Markarian SBGs. With increasing the starburst age, O stars begin to disappear. The ionizing-photon flux $N_{\rm ion}$ decreases, but the number of B stars and hence the FIR luminosity $L_{\rm FIR}$ still increase. Since hot large grains are progressively unimportant, the spectral index $\alpha (25, 60)$ becomes steep and approaches to the values observed in the Palomar SBGs.

\section{DISCUSSION}

While the Markarian SBGs are dominated by young objects ($t$ $\simeq$ 10 Myr), we argue that the Palomar SBGs with $L_{\rm FIR}$ $>$ 10$^{10}$ $L_{\odot}$ are dominated by old objects ($t$ $\simeq$ 10--100 Myr). The unbiased Palomar Survey provides a fair representation of the local galaxy population, where old SBGs are major constituents (30--40\% in number). 

The spectral index $\alpha (25, 60)$ lies between $-2$ and $-3$ in 90\% of the Palomar SBGs (Fig. 1$b$). This percentage agrees with our model. The starburst age and duration in classical SBGs are estimated as $t$ $\simeq$ $\tau$ $\simeq$ 10--20 Myr (Calzetti 1997; Leitherer 1998). For $\tau$ = 20 Myr, the values $\alpha (25, 60)$ = $-2$ and $-3$ are reproduced, respectively, at $t$ $\simeq$ 10 Myr and 100 Myr (Fig. 2$c$). The Markarian SBGs actually have the median value of $\alpha (25, 60)$ = $-2.06$.

The above estimation implies that old SBGs are $\sim$10 times more numerous than classical SBGs. The entire Palomar SBGs share 40\% of the surveyed galaxies (Ho et al. 1997a, b, c), while the Markarian SBGs share 3\% of the field galaxies (Balzano 1983). The Palomar SBGs with $L_{\rm FIR}$ $\le$ $10^{10}$ $L_{\odot}$ appear to be also dominated by old SBGs. The lower luminosity of their FIR emission is attributable to a weaker starburst or a greater age. Since, however, the {\it IRAS} fluxes of such objects could be contaminated by galaxy bulges and disks, we have not studied them.

Our model does not explain the difference between the Markarian and Palomar SBGs in the spectral index $\alpha (60, 100)$, which depends primarily on the temperature of warm large grains $T_{\rm WLG}$. The behavior of $\alpha (60, 100)$ is explained by the location of the warm large grains. A starburst region consists of star clusters. In young SBGs, nonionizing photons are absorbed by grains in the vicinity of the individual clusters. However, in old SBGs, supernova-driven shocks have pushed away those grains, and nonionizing photons are of lower energies, i.e., of longer wavelengths, and travel distances before absorbed by a grain. The absorption of nonionizing photons is due to grains which lie far from the clusters. Since the radiation field there is weak, the grain temperature is expected to be low.

The observed differences in the {\it IRAS} colors between the Palomar and Markarian SBGs are not attributable to a difference in the initial mass function. The initial mass function assumed in our calculation is consistent with most observations of star-forming and starburst regions, and thus expected to be universal (Leitherer 1998). Moreover, as stated above, the observed number of the Markarian SBGs requires the presence of numerous old SBGs.

The Palomar SBGs exhibit low H$\alpha$ luminosities, low [\ion{O}{3}] 5007\AA/H$\beta$ ratios, high [\ion{S}{2}] 6716\AA\ $+$ 6731\AA/H$\alpha$ ratios, and high [\ion{O}{1}] 6300\AA/H$\alpha$ ratios, as compared with classical SBGs. Ho et al. (1997c) already interpreted these properties in terms of old starbursts, where ionizing radiation is weak and of low excitation while supernova-driven shocks contribute to the [\ion{S}{2}] and [\ion{O}{1}] emissions. Nevertheless, our analysis based on FIR luminosities is essential to clarify the predominance of B stars in the Palomar SBGs. Since the beam size of the Palomar Survey is too small to compare its H$\alpha$ luminosity with the FIR luminosity, the spectral index $\alpha (25, 60)$ alone allows us to study the stellar population of the whole starburst region.

The Palomar SBGs prefer Hubble types equal to or later than Sbc (Ho et al. 1997b, c). We accordingly compare morphologies of the Palomar SBGs with those of the Markarian SBGs. Most of them are available in de Vaucouleurs et al. (1991). For the rest of the SBGs, we determine their morphologies on the basis of various references and optical images. The results are shown in Figure 1$d$. The Markarian SBGs have earlier morphologies than the Palomar SBGs (Ho et al. 1997c). Since the Markarian SBGs have optically prominent nuclei, they could be misclassified as earlier types. However, there is another possibility. Ho et al. (1997b) found that early-type galaxies from E to Sbc in the Palomar Survey preferentially harbor Seyferts and LINERs, classes of active galactic nuclei (AGNs). Young starbursts in early-type galaxies might tend to evolve into AGNs. The central engine of an AGN is a supermassive black hole, which is known to prefer a galaxy with a significant stellar bulge, corresponding to a Hubble type equal to or earlier than Sbc (Ferrarese \& Merritt 2000; Gebhardt et al. 2000).\footnote{
We expect that the Palomar and Markarian SBGs exhibit systematically different {\it IRAS} colors at each of the Hubble types. This expectation has been confirmed at least for the Hubble types around Sb, where the object numbers are sufficiently large. We also note that optical spectra of LINERs are close to those of shock-heated nebulae. Some of the Palomar LINERs might be old starbursts dominated by supernova-driven shocks (Larkin et al. 1998; Alonso-Herrero et al. 2000). Since, however, the galaxy morphologies are different, most of the Palomar LINERs are unlikely to have evolutionary connections with the Palomar SBGs.}

Vigorous star formation is found around many AGNs. Following Glass \& Moorwood (1985) and Mouri \& Taniguchi (1992), it is now established that these circumnuclear starbursts are older than those in classical SBGs (Moorwood 1996). Schmitt, Storchi-Bergmann, \& Cid Fernandes (1999) estimated the typical age of the circumnuclear starburst as $t$ $\simeq$ 100 Myr. Very young starbursts do not exist around AGNs. The  starburst occurs at radii of 100--1000 pc, while the fueling of the central engine occurs at radii of $\ll$ 1 pc. Since it takes time to transport gas from the circumnuclear region to the nucleus, there is a delay between the onset of the starburst and that of the AGN activity (Moorwood 1996; Schmitt et al. 1999). However, except for this difference, starbursts around AGNs are similar to those in SBGs, if the latter objects are represented by the Palomar SBGs.

Finally, we comment on objects in the {\it IRAS} Bright Galaxy Survey ($S_{60}$ $>$ 5.24 Jy; Soifer et al. 1989; see also Sanders \& Mirabel 1996). The majority of these {\it IRAS} galaxies at $L_{\rm FIR}$ $\le$ 10$^{11}$ $L_{\odot}$ are SBGs. With the data of Kim et al. (1995) and Veilleux et al. (1995), the median colors for 28 {\it IRAS} SBGs with $L_{\rm FIR}$ = 10$^{10}$--10$^{11}$ $L_{\odot}$ are estimated as $\alpha (25, 60)$ = $-2.30$ and $\alpha (60, 100)$ = $-1.16$. They are, respectively, close to the median values for the Palomar SBGs ($\alpha (25, 60)$ = $-2.34$) and the Markarian SBGs ($\alpha (60, 100)$ = $-1.01$). Thus the {\it IRAS} SBGs are at intermediate starburst ages. The ionizing radiation has become weak, but the nonionizing radiation is still strong and heats significantly the warm large grains. The space density of the {\it IRAS} SBGs appears to be greater than that of the Markarian SBGs.

\acknowledgments
The authors are grateful to the referee for useful comments. This work was supported in part by Grant-in-Aids for the Scientific Research (Nos. 10044052, and 10304013) of the Japanese Ministry of Education, Culture, Sports, and Science.

\clearpage

\figcaption[Fig1]{Statistical properties of Palomar SBGs ({\it filled areas}) and Markarian SBGs ({\it open areas}): ($a$) FIR luminosity $L_{\rm FIR}$; ($b$) spectral index $\alpha (25, 60)$; ($c$) spectral index $\alpha (60,100)$; and ($d$) Hubble type. Upper limits are indicated by wedges. In ($b$)--($d$), we analyze exactly the same objects with $L_{\rm FIR}$ $>$ 10$^{10}$ $L_{\odot}$. In ($d$), the bins along the abscissa have the following means: ``Uc'' = unclassified, ``E'' = E, ``S0'' = S0, ``Sa'' = S0/a--Sab, ``Sb'' = Sb--Sbc, ``Sc'' = Sc--Scd, ``Sd'' = Sd--Sdm, ``Sm'' = Sm--Im, ``I0'' = I0, and ``Pec'' = Pec $+$ S pec. The objects classified by the authors are Mrk 316, Mrk 717 (``E''), Mrk 1002 (``S0''), Mrk 739E (``Sa''), Mrk 21, Mrk 353, Mrk 489, Mrk 684, Mrk 764 (``Sb''), Mrk 307, Mrk 373E, NGC 7798 (``Sc''), Mrk 496, Mrk 809 (``Sd''), Mrk 413 (``Sm''), Mrk 789 (``I0''), Mrk 480, NGC 812, NGC 2342, and UGC 3714 (``Pec''). There are three unclassified objects, Mrk 357, Mrk 420, and Mrk 719, which appear to be early-type galaxies. The original Palomar and Markarian SBGs consist of 206 and 102 objects. For some of them, the {\it IRAS} data are not available. The number of the Palomar and Markarian SBGs used in the analyses are ($a$) 198 and 84, and ($b$)--($d$) 56 and 44. Five objects in ($a$) and three objects in ($b$)--($d$) are common to the two samples. The median values for the Palomar and Markarian SBGs are ($a$) 9.60 and 10.01 in $\log (L_{\rm FIR}/L_{\odot})$, ($b$) $-2.34$ and $-2.06$, and ($c$) $-1.62$ and $-1.01$. Here we have used the Kaplan-Meier estimator for data with upper limits (Feigelson \& Nelson 1985). \label{Fig.1}}

\figcaption[Fig2]{Results for model calculations: ($a$) ionizing-photon flux $N_{\rm ion}$, ($b$) FIR luminosity $L_{\rm FIR}$, and ($c$) spectral index $\alpha (25, 60)$. The abscissa is the starburst age $t$. The star-formation rate is assumed to decay exponentially as $\exp (-t/ \tau )$, where the starburst duration $\tau$ is set to be 1, 2, 5, 10, 20, or 50 Myr. The results for continuous star formation are also shown ($\tau$ = $\infty$). In ($a$) and ($b$), the $N_{\rm ion}$ and $L_{\rm FIR}$ values are normalized by their maximum values. In ($c$), we show the effects of changing the temperature of warm large grains $T_{\rm WLG}$ between 30 and 40 K at $t$ = 1 and 300 Myr. \label{Fig.2}}

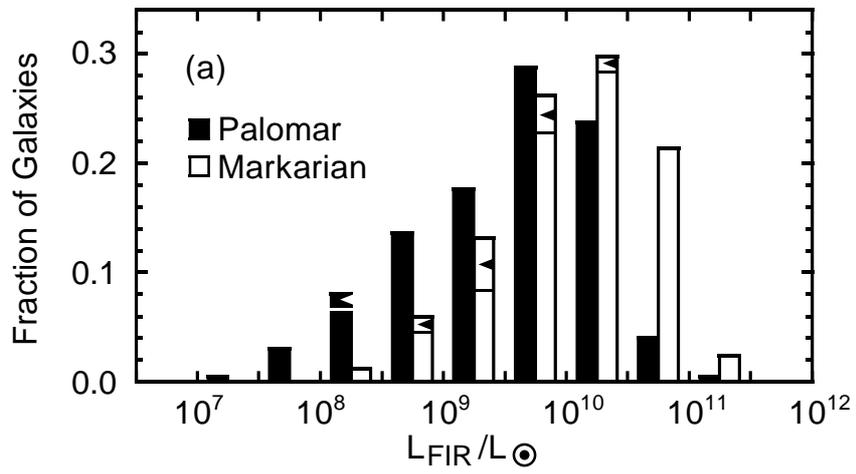
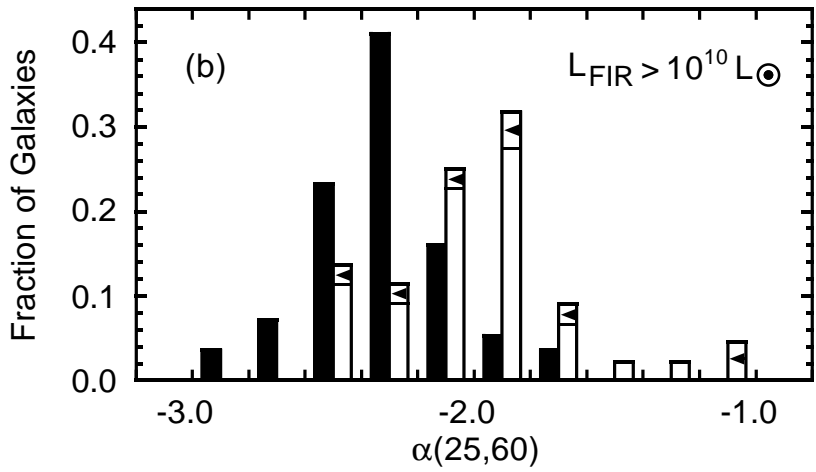
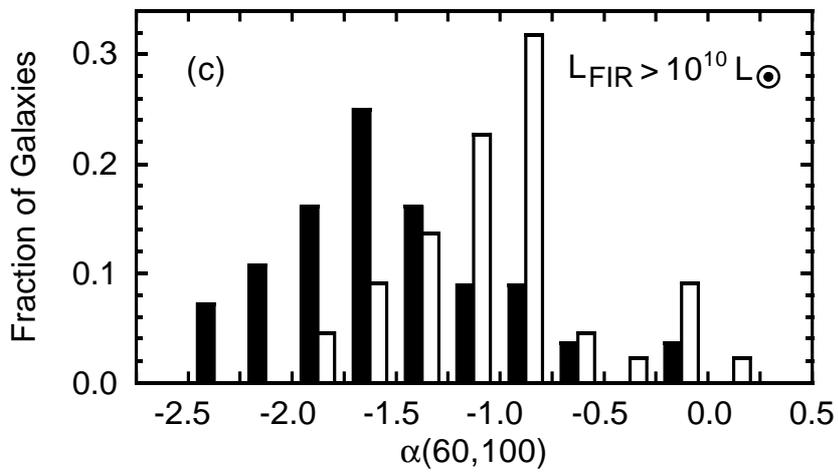
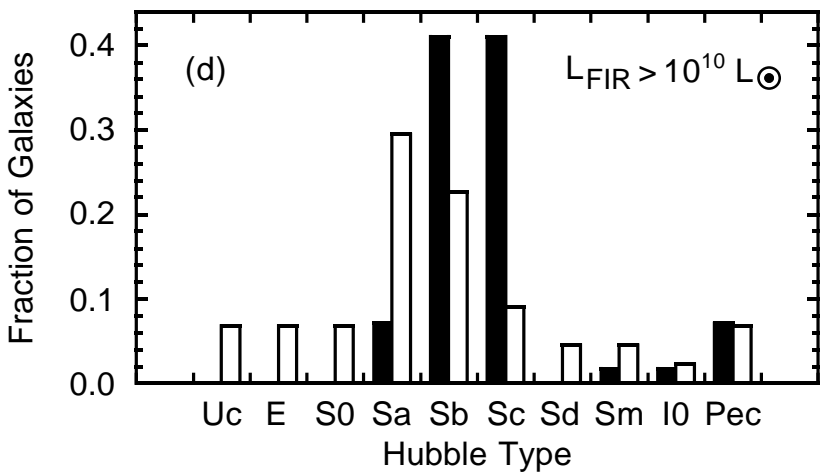
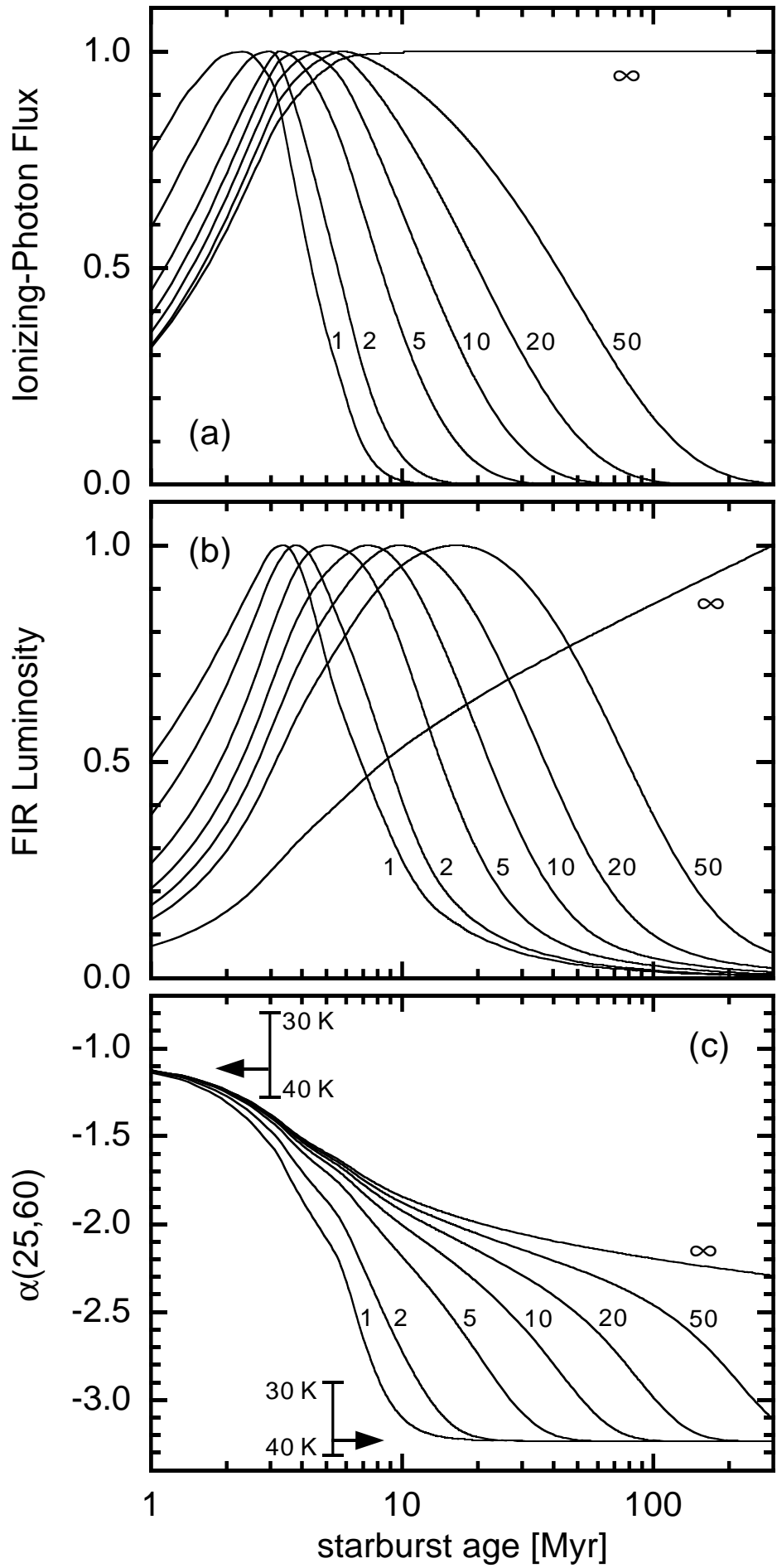
\end{document}